\documentclass[twoside,final]{pro12}
\usepackage[ansinew]{inputenc}
\usepackage{amsmath}
\usepackage[dvips]{graphicx} % use graphics

\head{G. Fischer et al.} {Saturn lightning}

\begin{document}

\title{OVERVIEW OF SATURN LIGHTNING OBSERVATIONS}
\author{G. Fischer\adress{\textsl{Space Research Institute, Austrian
Academy of Sciences, Graz, Austria}}$\,$, U. A.
Dyudina\adress{\textsl{California Institute of Technology, Pasadena,
USA}}$\,$, W. S. Kurth\adress{\textsl{Dept. of Physics and
Astronomy, The University of Iowa, Iowa City, USA}}$\,$, D. A.
Gurnett$^\ddagger$,\\ P. Zarka\adress{\textsl{LESIA, Observatoire de
Paris--Meudon, Meudon Cedex, France}}, T. Barry\adress{\textsl{
Broken Hill (Barry) and Murrumbateman (Wesley), Australia}}$\,$, M.
Delcroix\adress{\textsl{Commission des observations plan\'{e}taires,
Soci\'{e}t\'{e} Astronomique de France}}$\,$, C. Go\adress{\textsl{
University of San Carlos, Philippines}}$\,$, D.
Peach\adress{\textsl{British Astronomical Association, United
Kingdom}}$\,$,\\ R. Vandebergh\adress{\textsl{Wittem, The
Netherlands}}$\,$, and A. Wesley$^\P$}

% Here the third author has the same affiliation as the second author.
% Emails addresses are not required, as we will make a separate email list of all
% participants at the end of the proceedings book (see PRE VI book).

\maketitle

\begin{abstract}
% Enter the text of your abstract here
The lightning activity in Saturn's atmosphere has been monitored by
Cassini for more than six years. The continuous observations of the
radio signatures called SEDs (Saturn Electrostatic Discharges)
combine favorably with imaging observations of related cloud
features as well as direct observations of flash--illuminated cloud
tops. The Cassini RPWS (Radio and Plasma Wave Science) instrument
and ISS (Imaging Science Subsystem) in orbit around Saturn also
received ground--based support: The intense SED radio waves were
also detected by the giant UTR--2 radio telescope, and committed
amateurs observed SED--related white spots with their backyard
optical telescopes. Furthermore, the Cassini VIMS (Visual and
Infrared Mapping Spectrometer) and CIRS (Composite Infrared
Spectrometer) instruments have provided some information on chemical
constituents possibly created by the lightning discharges and
transported upward to Saturn's upper atmosphere by vertical
convection. In this paper we summarize the main results on Saturn
lightning provided by this multi--instrumental approach and compare
Saturn lightning to lightning on Jupiter and Earth.
\end{abstract}

\section{Radio observations of SEDs by Cassini RPWS}

\begin{figure}[ht]
\centering
\includegraphics[width=1\textwidth]{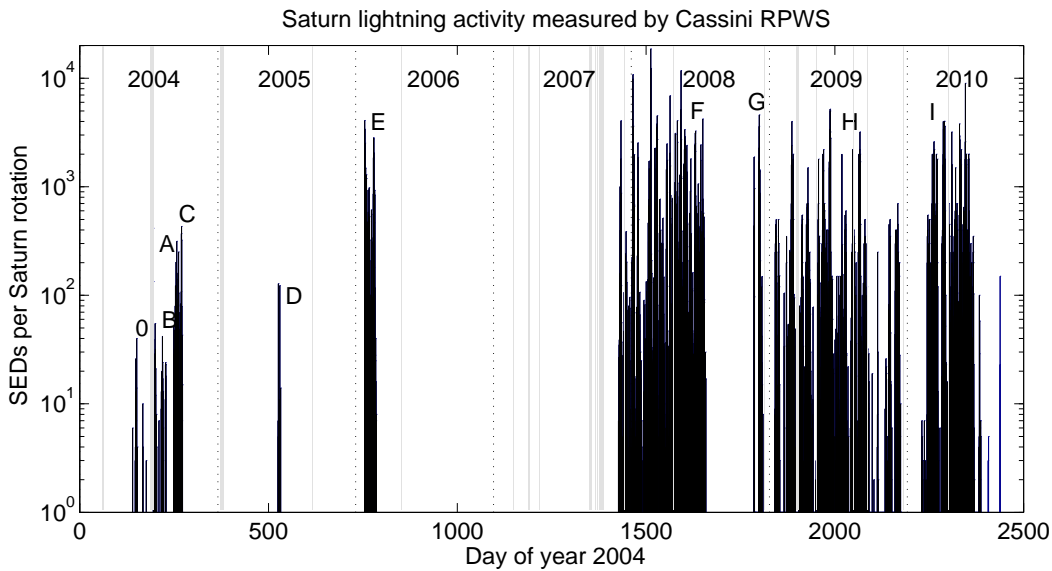}
\caption{Number of SEDs recorded by Cassini/RPWS as a function of
time. The gray background denotes data gaps, and the SED numbers
from 2009 and later are estimates.} \label{ffig1} \centering
\end{figure}

Saturn Electrostatic Discharges (SEDs) are short and strong radio
bursts that were initially detected by the radio instrument
on--board Voyager~1 near Saturn [Warwick et al., 1981]. The SEDs
were again detected by the Cassini RPWS instrument around Saturn
Orbit Insertion [Fischer et al., 2006], and since then there has
been practically continuous monitoring. Until the time of this
writing the RPWS instrument has recorded 10 storms of Saturn
lightning (named 0, A, B, C, D, E, F, G, H, and I) which are
displayed in Figure~\ref{ffig1} showing the number of SEDs per
Saturn rotation as a function of time. Table~1 lists these SED
storms giving their name, their start and stop days, the number of
SEDs and episodes, and the re--occurrence period of the episodes. It
is immediately obvious from Figure~1 that the SED activity in the
last three years was particularly strong, whereas in the first three
years of the Cassini mission the lightning activity was much more
sparse. There was even an interval of 21 months from February 2006
until November 2007 with no SED activity. The periodic occurrence of
great white spots with each Saturnian year [Sanchez--Lavega et al.,
1991] suggests that the atmosphere of Saturn undergoes seasonal
changes. It could be that the strong SED activity of the last years
is linked to Saturn's equinox (August 11, 2009). Until the end of
the Cassini mission in September 2017 the spacecraft will have
stayed in Saturn's orbit for $\sim 45$\% of a Saturnian year. This
and future ground--based optical and radio observations should
enable the identification of a possible seasonality of Saturn
lightning storms.

\begin{table}[h]
\caption{Saturn lightning storms recorded by Cassini/RPWS. The SED
numbers give the number of detected SEDs plus the number of SED
episodes (Saturn rotations during which SEDs were present). For the
last two storms H and I the exact numbers still have to be
determined (TBD).}\label{ftable1}
\begin{center}
\begin{tabular}{|c|c|c|c|} % this is to center (c) all numbers
% with vertical lines in between, use l or r for flush left or right
\hline % this makes a horizontal line
Name & Start and stop date & SED numbers & Period\\
\hline
0 & May 26--31, 2004 & 100 in 8 & 10h35min ($\pm6$min)\\
A & July 13--27, 2004 & 800 in 15 & 10h43min ($\pm3$min)\\
B & August 3--15, 2004 & 300 in 16 & 10h40min ($\pm3$min)\\
C & Sept. 4--28, 2004 & 4,200 in 49 & 10h40min ($\pm1$min)\\
D & June 8--15, 2005 & 300 in 6 & 10h10min?\\
E & Jan. 23 -- Feb. 23, 2006 & 43,400 in 71 & 10h40min ($\pm0.4$min)\\
F & Nov. 27, 2007 -- July 15, 2008 & 282,300 in 439 & $\sim$10h40min\\
G & Nov. 19 -- Dec. 11, 2008 & 22,000 in 22 & $\sim$10h40min\\
H & Jan. 14 -- Dec. 13, 2009 & TBD in $\sim 470$ & $\sim$10h40min\\
I & Feb. 7 -- July 14, 2010 & TBD in $\sim 270$ & $\sim$10h40min\\
\hline
\end{tabular}
\end{center}
\end{table}

\section{Ground--based radio observations of SEDs}

SED signals are at least $10^4$ times stronger than the radio
signals of terrestrial lightning in the frequency range of a few MHz
[Fischer et al., 2006]. Several unsuccessful attempts were made to
detect SEDs with ground--based radio telescopes. The negative
results were either due to insufficient sensitivity or the intrinsic
variability of SED occurrence [Zarka et al., 2008]. Finally, the
high intensity of the SEDs and the real time information of Cassini
about on--going lightning activity enabled the first convincing
detection of SEDs by the large ground--based radio telescope UTR--2
in the Ukraine in early 2006 [Konovalenko et al., 2006]. For a
detailed description we refer the reader to the article by
Grie{\ss}meier et al. [2011] in this issue, and we only mention that
SEDs are a good observational target for large radio telescopes like
UTR--2 or LOFAR.

\section{Direct optical observations of Saturn lightning flashes}

\begin{figure}[ht]
\centering
\includegraphics[width=1\textwidth]{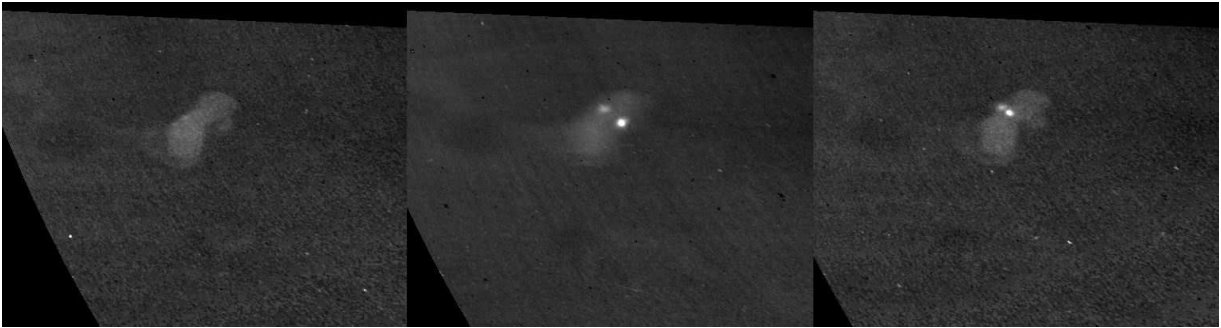}
\caption{Sequence of three images by Cassini ISS from November 30,
2009, showing the SED storm cloud on the night side at $35^{\circ}$
south with lightning flashes (\copyright NASA).} \label{ffig2}
\centering
\end{figure}

The first direct optical observation of Saturn lightning by Cassini
ISS occurred on August 17, 2009, and was recently published by
Dyudina et al. [2010]. The reduced ring shine around equinox allowed
the cameras to detect illuminated cloud tops on Saturn's night side.
The size of the spots of a few hundred kilometers allowed the
determination of the SED source depth which should be located
125--250~km below the cloud tops, most likely in the water cloud
layer. From the brightness of the spots Dyudina et al. [2010] also
inferred an optical energy of $\sim 10^9$~J, which corresponds to
the total energy of a terrestrial lightning flash. In
Figure~\ref{ffig2} we show another direct flash observation from
November 30, 2009. As in August around equinox, the storm system was
located at a planetocentric latitude of $35^{\circ}$ in the southern
hemisphere, which at that time was still immersed in relative
darkness (northern side of the rings illuminated). The storm clouds
in Figure~\ref{ffig2} and in all the other images of this paper
should be due to moist convective storms that develop vertical
convective plumes that overshoot the outermost ammonia (NH$_3$)
cloud layer [Stoker, 1986; Hueso and Sanchez--Lavega, 2004].

\section{Optical observations of storm clouds}

\begin{figure}[ht]
\centering
\includegraphics[width=0.49\textwidth]{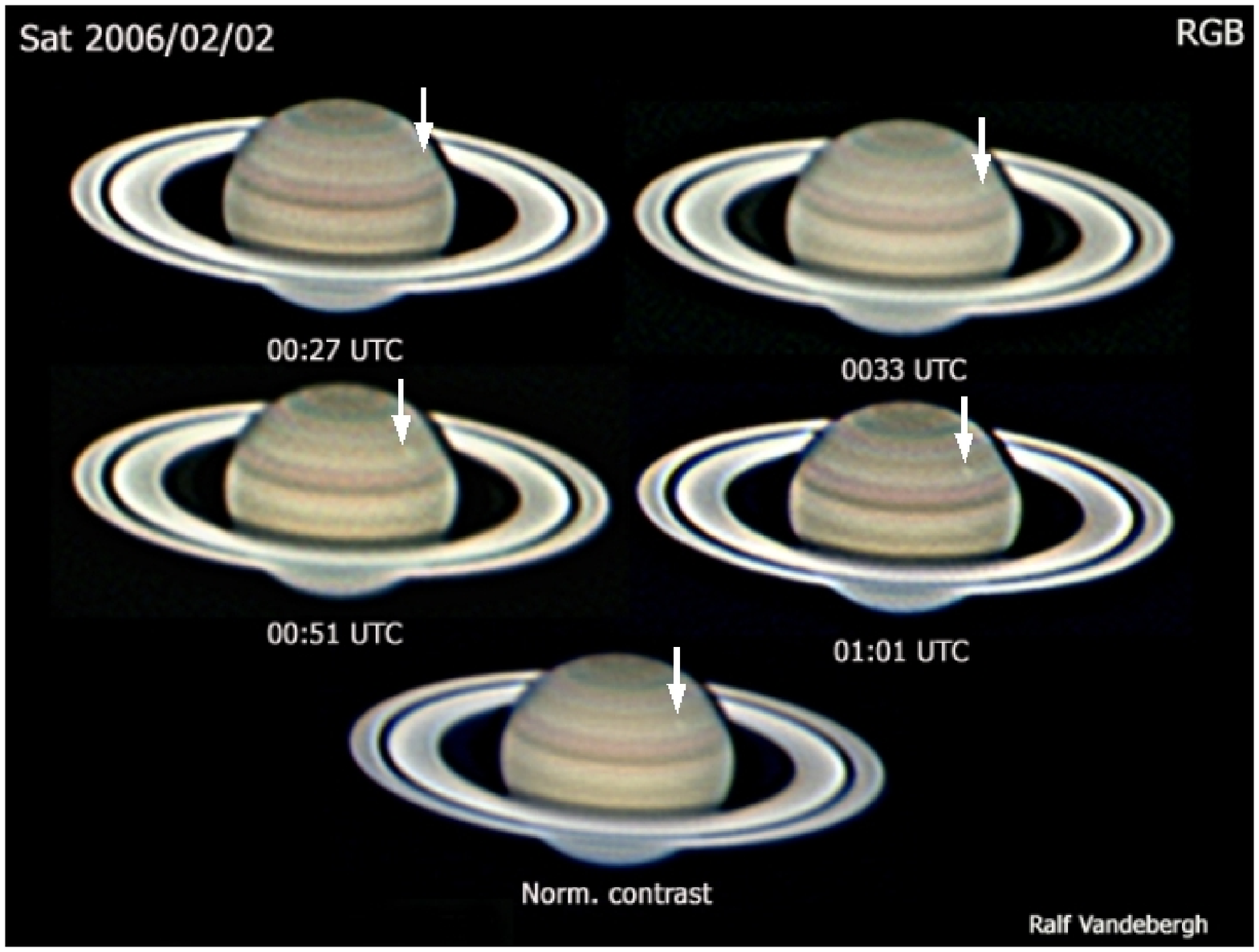}
\includegraphics[width=0.49\textwidth]{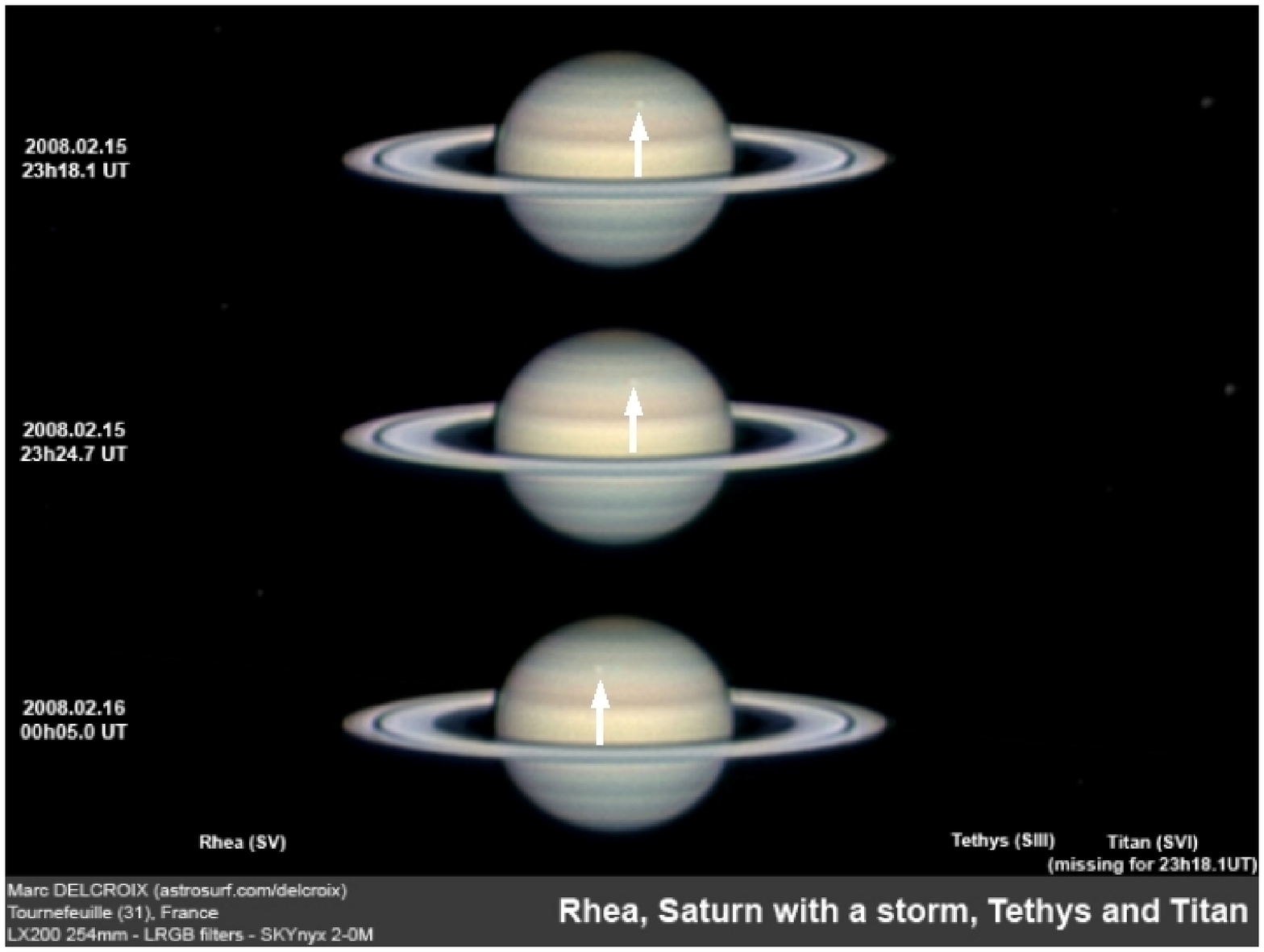}
\caption{Saturn amateur images showing white spots around
$35^{\circ}$ south during SED storms E and F, respectively. The
sequence of 5 Saturn images on the left side was taken on February
2, 2006, by Ralf Vandebergh from the Netherlands. The sequence of 3
images on the right side was taken on February 15, 2008, by Marc
Delcroix from France.} \label{ffig3} \centering
\end{figure}

% Enter the text of your third section
The first link between SEDs and optical observations of storm clouds
was found in 2004, when the so--called "dragon storm" was imaged by
Cassini ISS together with SED detections by RPWS [Porco et al.,
2005; Fischer et al., 2006]. These combined observations revealed
consistent longitudes and longitudinal drift rates of the cloud
feature with respect to the SED source. The white storm clouds were
also found to be brighter when the SED rates were high [Dyudina et
al., 2007; Fischer et al., 2007]. The "dragon storm" was located at
a planetocentric latitude of $35^{\circ}$ south. This region is
nicknamed "storm alley", and all SED storms shown in this section
were observed there.\\
The involvement of amateur astronomers in the observation of
Saturn's storms started with the one month long SED storm E in
January/February 2006. It was the first long--lasting SED storm
observed by Cassini when Saturn was far from solar conjunction and
thus high in the sky for ground--based observers. In the amateur's
images the SED storms turned out to be detectable relatively easily,
and they appear as bright white spots. We marked these spots with
white arrows in the following figures since the loss in image
quality by size reduction and printing makes some of them hard to
discern. The left side of Figure~\ref{ffig3} shows a sequence of 5
Saturn images taken by Ralf Vandebergh on February 2, 2006. It can
be seen how a white spot at a latitude around $35^{\circ}$ south
progressively comes into view from the right side, i.e. the western
horizon since south is pointed upwards. Similarly, the sequence of 3
images on the right side of Figure~\ref{ffig3} shows a storm system
around the central meridian. It was taken nearly two years later by
Marc Delcroix during SED storm F, and the tilt of the rings has
decreased significantly.

\begin{figure}[ht]
\centering
\includegraphics[width=0.49\textwidth]{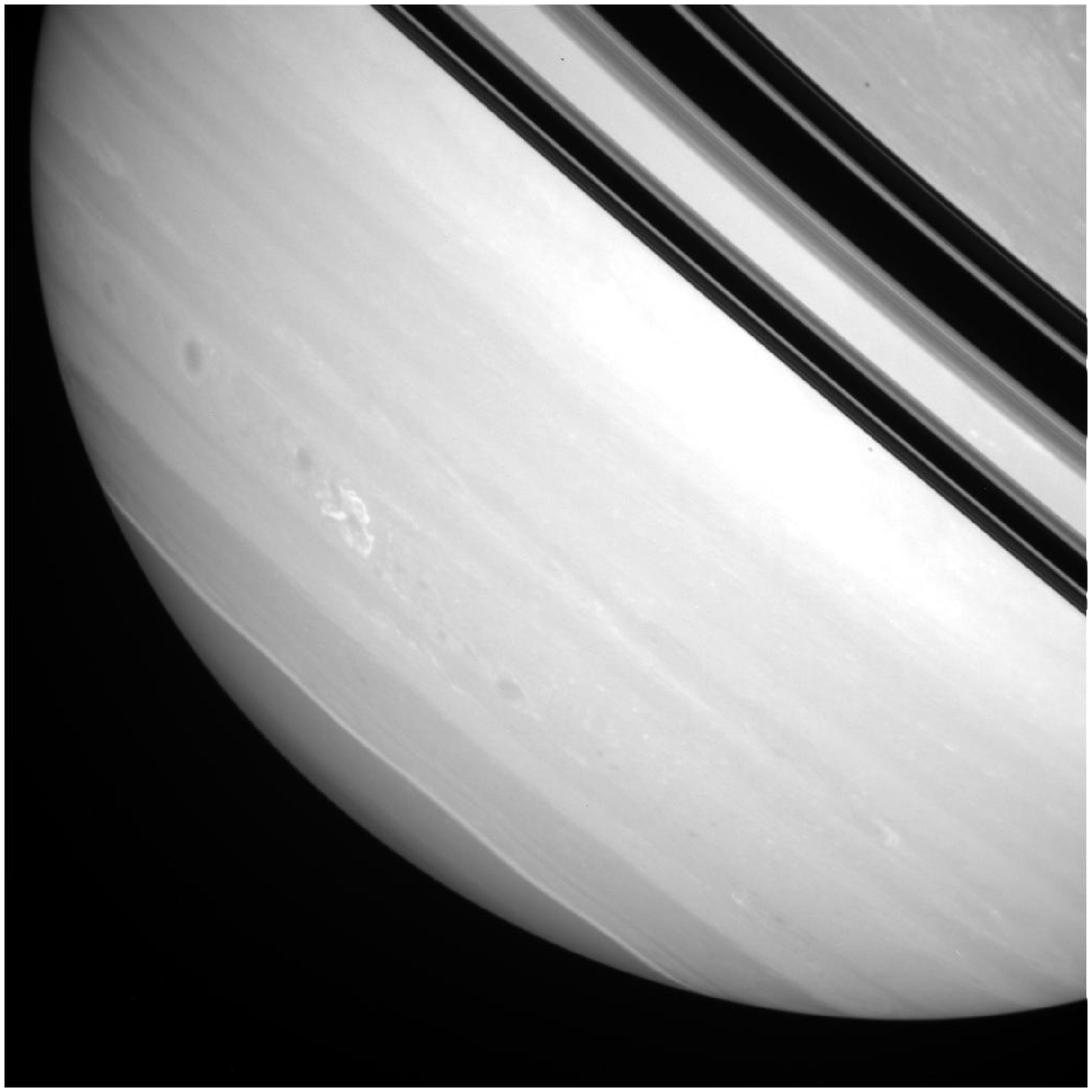}
\includegraphics[width=0.49\textwidth]{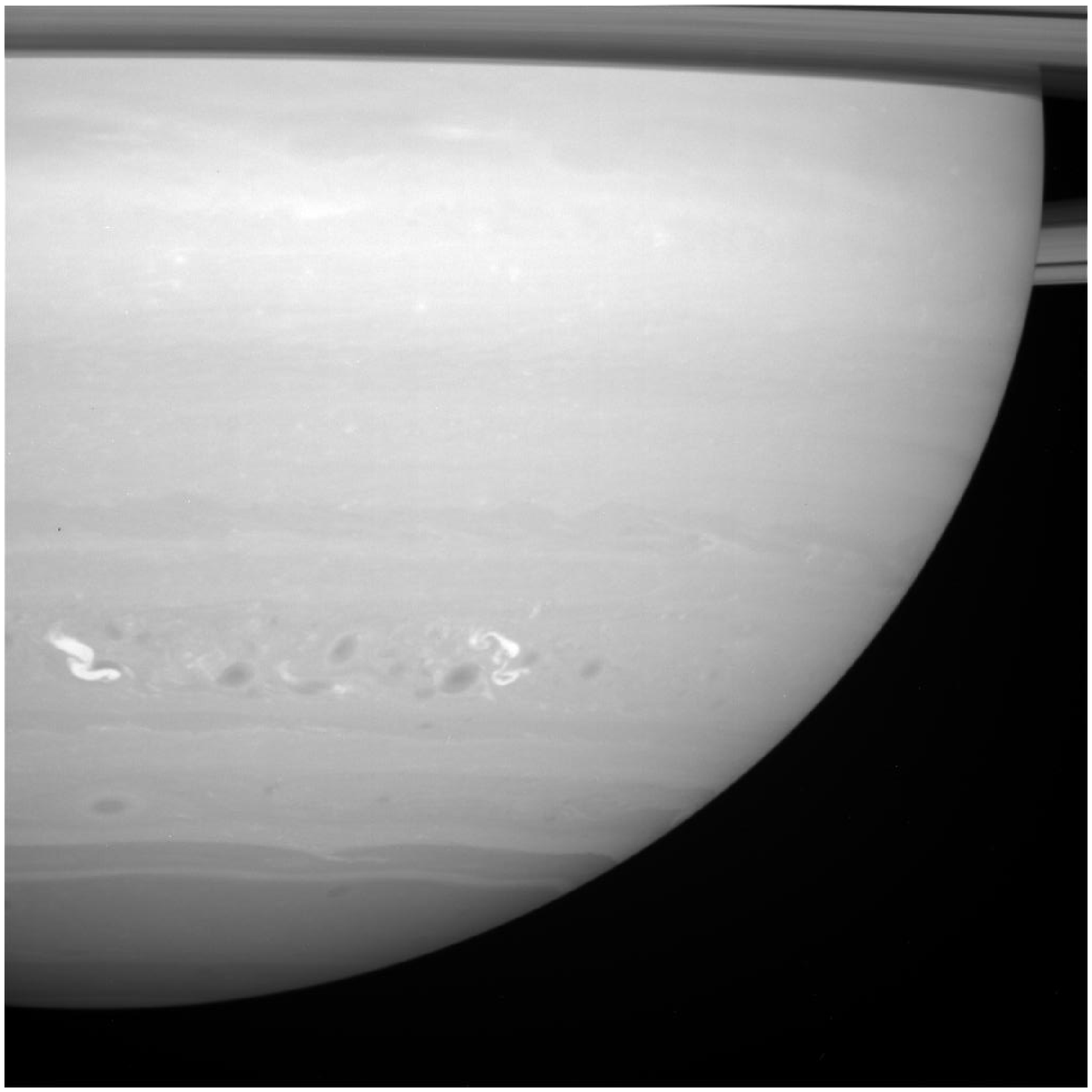}
\caption{Cassini ISS images from March 4, 2008 (left), and June 18,
2008 (right side). The left image shows only one bright storm
system, whereas there are two in the right image (\copyright NASA).}
\label{ffig4} \centering
\end{figure}

\begin{figure}[ht]
\centering
\includegraphics[width=0.519\textwidth]{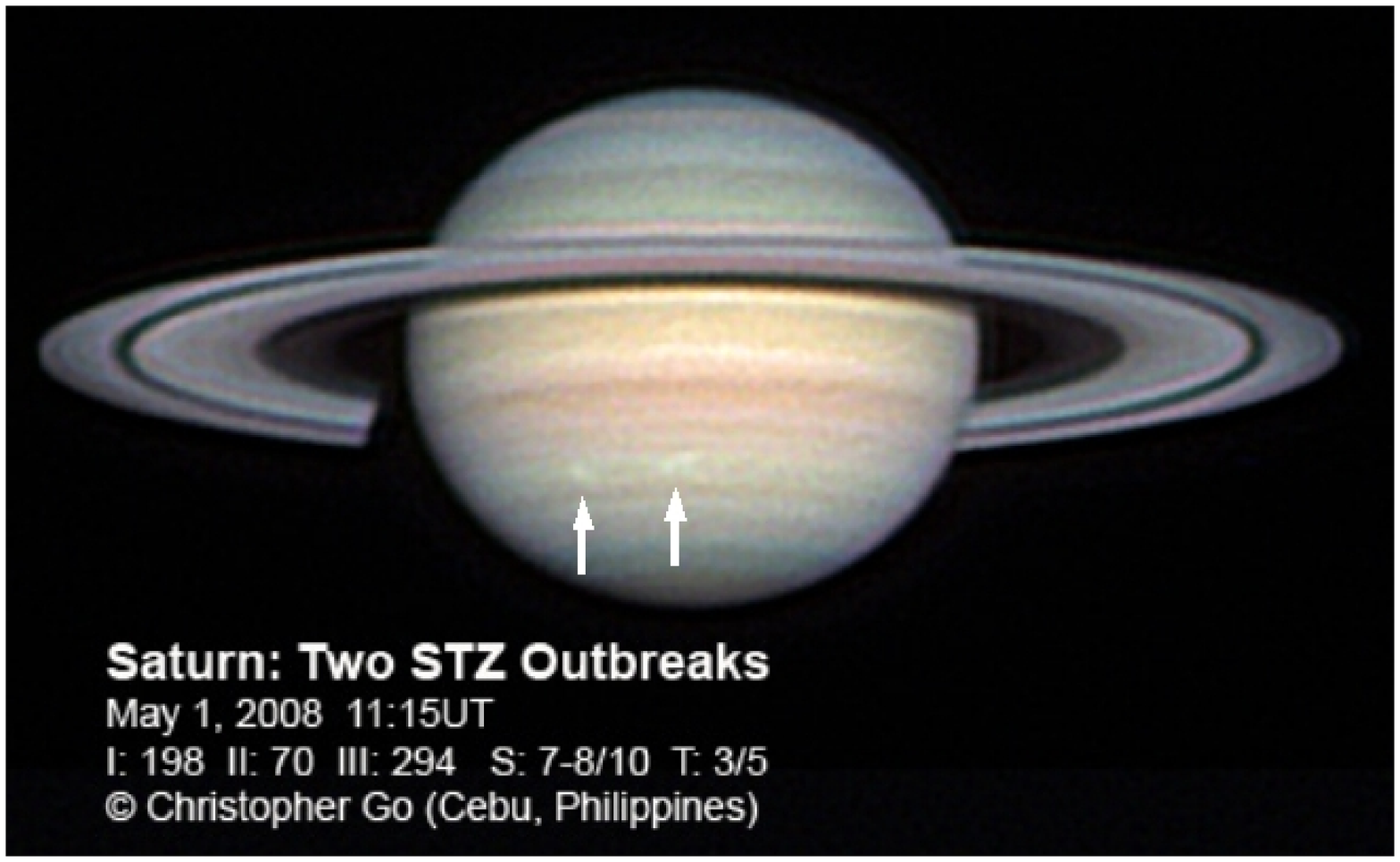}
\includegraphics[width=0.461\textwidth]{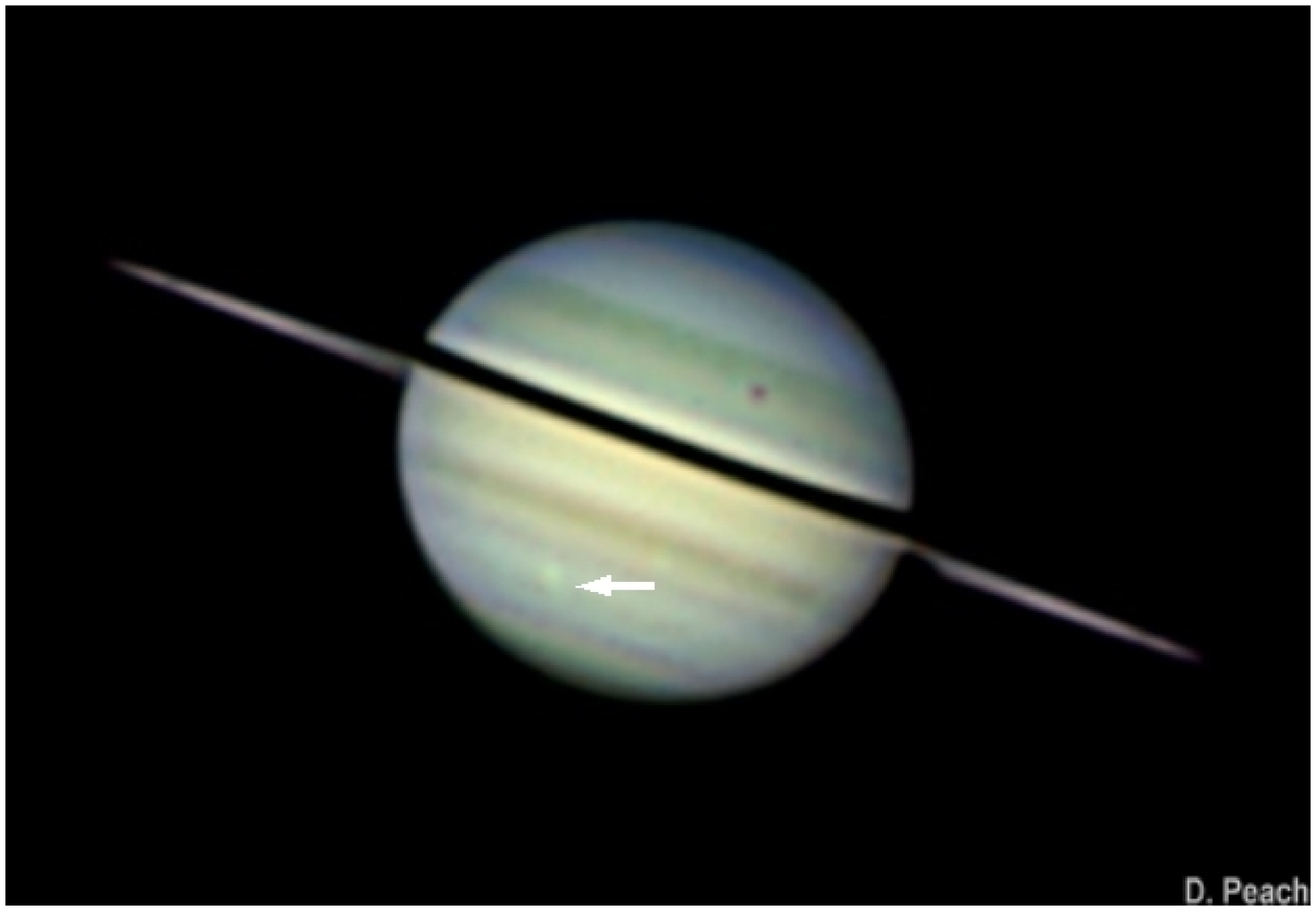}
\caption{Saturn amateur images showing white spots around
$35^{\circ}$ south during SED storm F and G, respectively. The image
on the left side shows two storms and was taken on May 1, 2008, by
Christopher Go from the Philippines. The right image was taken on
December 7, 2008, by Damian Peach from the United Kingdom. South is
downward in both images.} \label{ffig5} \centering
\end{figure}

Figure~4 shows two images from Cassini ISS taken during the 7.5
months long SED storm F which consisted of two phases. In the first
phase from the end of November 2007 until early March 2008 only one
storm system was present as imaged by Delcroix (right side of
Figure~\ref{ffig3}) and by ISS (left side of Figure~\ref{ffig4}).
From March 4--10, 2008, the RPWS instrument recorded uninterrupted
SED activity for 9 consecutive Saturn rotations. This was a unique
event since normally the SED activity is organized in episodes with
periodic gaps when the storm is located beyond the radio horizon on
the far side of the planet as seen from Cassini. The only
explanation is that there was at least a second storm system
present, and investigations by Delcroix and Fischer [2010] revealed
that there were most likely even three storm systems at the same
time in March 2008. After those 9 Saturn rotations the typical
episodic behavior continued, but the longitude range at which SED
activity was recorded seemed somewhat larger than usual. The reason
for this was soon seen in the images, and this constitutes the
second phase of SED storm F. A relatively persistent second storm
system had developed which was located at the same latitude, but the
two storms were separated by about $30^{\circ}$ in longitude. The
right side of Figure~\ref{ffig4} clearly shows the two white clouds
imaged by Cassini ISS on June 18, 2008. Ground--based observers even
imaged them earlier, and the left side of Figure~\ref{ffig5} shows
an image with the two storms taken by Christopher Go on May 1, 2008.

\begin{figure}[ht]
\centering
\includegraphics[width=0.474\textwidth]{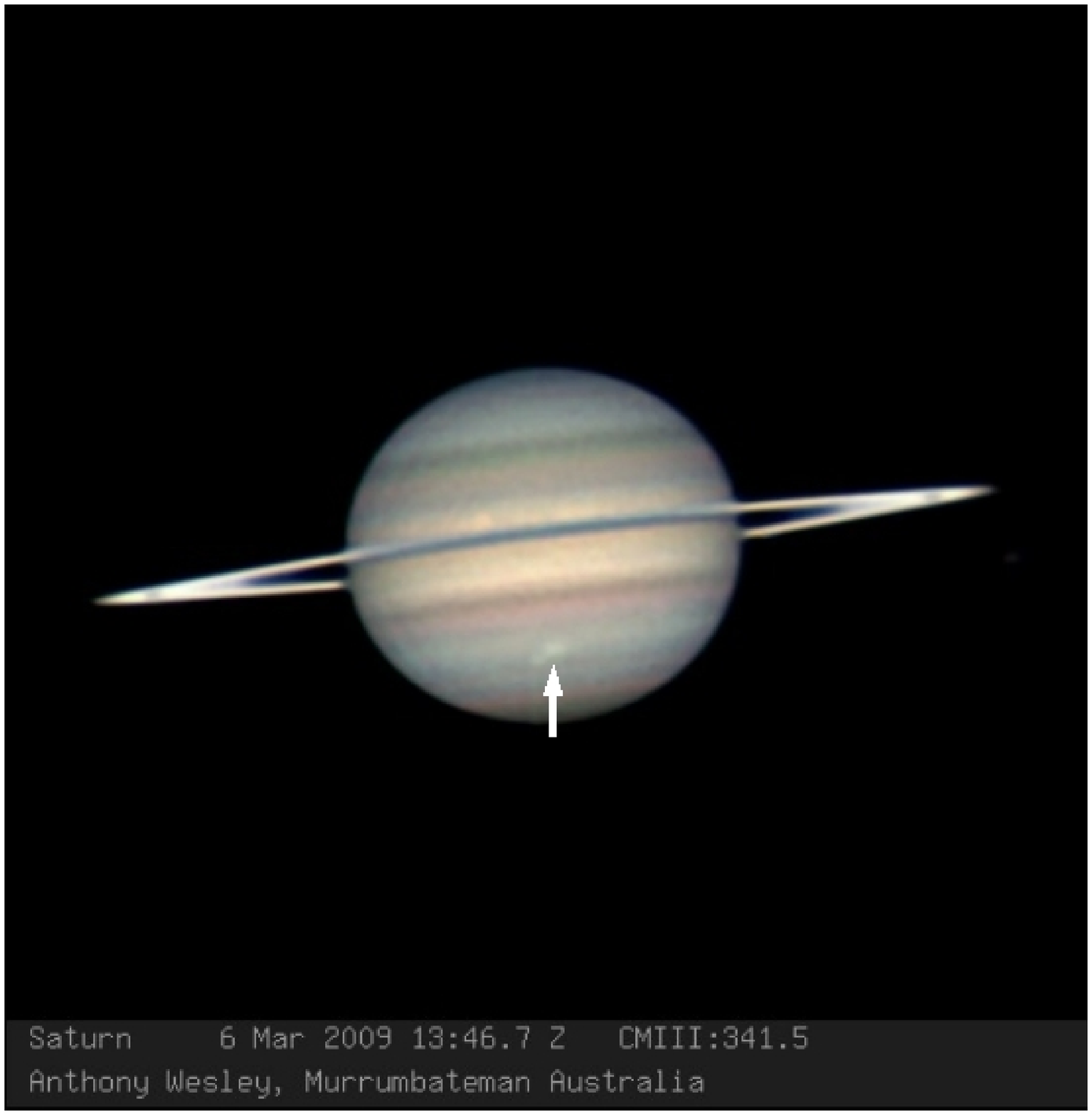}
\includegraphics[width=0.506\textwidth]{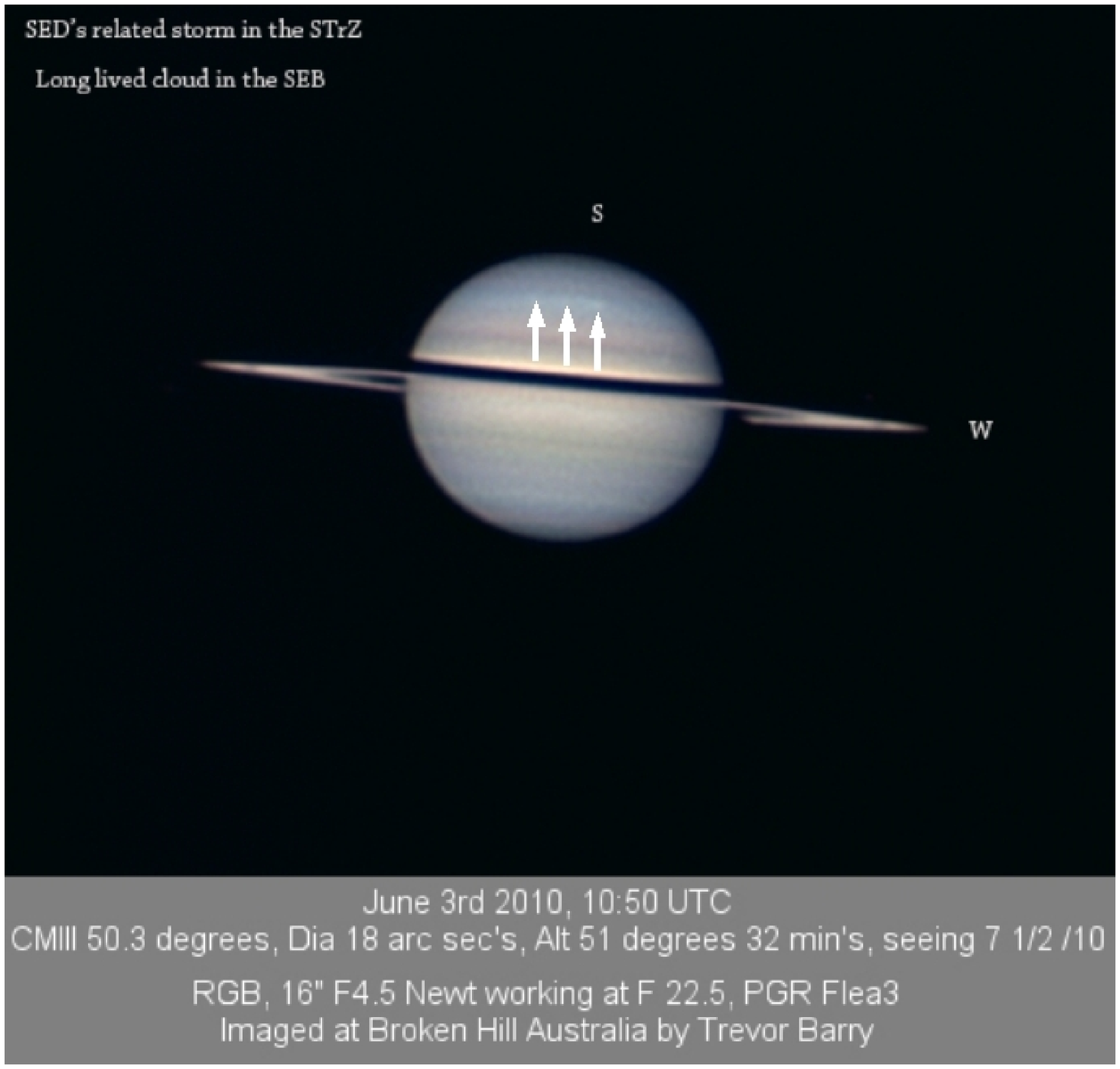}
\caption{Saturn amateur images by the two Australians Anthony Wesley
(left side, taken on March 6, 2009) and Trevor Barry (right side,
taken on June 3, 2010). They show the white spots in Saturn's storm
alley at $35^{\circ}$ south during SED storm H and I, respectively.
South is downward in the image by Wesley, but upward in the image by
Barry.} \label{ffig6} \centering
\end{figure}

The next SED storm G lasted only for about three weeks and took
place in November/December 2008. Again, one white spot was seen at
$35^{\circ}$ south as shown on the right side of Figure~\ref{ffig5}
in an image by Damian Peach (dark spot is the shadow of Dione). SED
storm G was a kind of precursor to the next SED storm H that lasted
from mid--January until mid--December 2009. With a duration of 11
months this SED storm is the longest ever observed lightning storm
in our solar system to date. The left image of Figure~\ref{ffig6}
was taken by Anthony Wesley on March 6, 2009, with a white spot
clearly visible around $35^{\circ}$ south in the storm alley.
Another interesting feature of this image is the unusual brightness
of Saturn's rings. This is the so--called Seeliger effect which
enhances the brightness of reflective objects that are in opposition
to the Sun. The image was taken just two days before the Earth moved
through the line between Saturn and the Sun. The last image we show
on the right side of Figure~\ref{ffig6} was taken by Trevor Barry on
June 3, 2010. Saturn had passed through equinox on August 11, 2009,
and the northern side of the rings is illuminated in this image
where south is pointing upwards. SED storm I of 2010 lasted for
about 5 months, and like the other storms it started with a single
white spot at $35^{\circ}$ south. However, several amateur images
revealed a stretching of the storm in longitude with time. A few
weeks before it faded away, there were actually three bright regions
(marked by three white arrows) next to each other as it can be seen
in the image by Barry.

In summary, for 8 of the 10 SED storms Cassini ISS and/or
ground--based observers detected an associated cloud feature in the
storm alley at a latitude of $35^{\circ}$ south. For only two SED
storms (namely 0 and D, see Table~\ref{ftable1}) no associated storm
cloud could be found, either due to the briefness and weakness of
the SED storm or missing image coverage. These two storms occurred
too close to solar conjunction to be observable from Earth. The
ground--based optical observations of the amateurs have become
increasingly important for Saturn lightning studies since Cassini
ISS has lots of other observational targets. The major advantage of
the RPWS instrument is the practically constant coverage, but the
radio observations can only reveal an approximate location of the
storms. The Cassini cameras can provide a very accurate positioning,
but sometimes lack coverage. This is where the amateurs can help and
during Saturn apparitions their coverage is usually good and
sufficient for longitudinal drift rate measurements and observations
of the storm shape evolution [Delcroix and Fischer, 2010].
Currently, there are more than 200 amateurs who have registered at
the website of the PVOL, the Planetary Virtual Observatory and
Laboratory at http://www.pvol.ehu.es/. This website was set up by
the Planetary Sciences Group of the Universidad del País Vasco in
Bilbao, Spain [Hueso et al., 2010]. Amateurs can post their
planetary images there which can then be viewed by everyone. We
inspected the catalog of Saturn images which goes back until
2001/2002. We found similar white spots in Saturn's southern storm
alley during the Saturn apparitions of 2002/2003 and 2003/2004, but
none during 2001/2002. It is very likely that these prominent white
spots were also related to SED storms that could not be detected by
the RPWS instrument due to the large distance of Cassini to Saturn
at that time.

\section{Other observations by Cassini VIMS and CIRS}

Bar--Nun [1975] predicted the existence of lighting on Jupiter prior
to its detection by the Voyagers based on the presence of
considerable acetylene (C$_2$H$_2$) concentration. He first
suggested that lightning may be the source of non--equilibrium
concentration of various compounds in the atmospheres of the gas
giants, and so new Cassini observations at Saturn are important to
investigate lightning--induced atmospheric chemistry. The Cassini
instruments VIMS and CIRS can observe Saturn's atmosphere at
infrared wavelength and therefore reveal some chemical information.
Spectroscopic observations by VIMS identified carbon soot particles
attached to ammonia (NH$_3$) and ammonium hydrosulfide (NH$_4$SH)
condensates in dark spots close to Saturn's storm clouds [Baines et
al., 2009]. The dark spots develop out of active lightning storms
and usually drift to the west [Dyudina et al., 2007]. They are
brought up to higher altitudes around 0.4~bar by vertical
convection, and they can be seen at the same latitude like the white
spots in both ISS images of Figure~\ref{ffig4}. Their dark color
might come from the soot that is thought to be created by
dissociation from methane (CH$_4$) in the high temperature lightning
channel. In a news feature from April 29, 2010, published on the
Cassini webpage, the CIRS team announced that it had found an
enhanced level of phosphine (PH$_3$) potentially brought up to
higher levels by convection. They also suspected that the lightning
storm is similar to an ammonia--ice blizzard that is going on a few
hundred kilometers below the tropopause.
%\begin{figure}[ht]
%\centering
%\includegraphics[width=0.9\textwidth,angle=-90]{figure1.eps}
% Put figure1.eps in the same directory as your manuscript LATEX file
% and use the width to determine the width of the figure with regard
% to the text (here 95% of text width, maximum is 100%, set 0.95 to 1).
% Use the option angle to rotate the figure.
%\caption{Text of figure caption}
%\centering
%\end{figure}
% To put two figure-files next to each other use two \epsfig commands:
%\includegraphics[width=0.45\textwidth]{figure1_1.eps}
%\includegraphics[width=0.45\textwidth]{figure1_2.eps}
% by reducing the textwidth to <0.5; it is counted as one figure.

\section{Comparison of lightning on Saturn, Jupiter, and Earth}

On Saturn lightning is intermittent, and usually there is only one
large storm which can last for several months. On Jupiter there are
typically a few storms at the same time, but they are about half the
size of the Saturnian ones and only last for a few days. Terrestrial
thunderstorms have an average duration of half an hour and an
average size of 25~km, but there are about 2000 thunderstorms in
progress at any given moment [Rakov and Uman, 2003]. These and other
characteristics are compared in Table~\ref{ftable2}.

\begin{table}[h]
\caption{Comparison of lightning on Saturn, Jupiter, and Earth. The
following references were used: $^a$Dyudina et al. [2007],
$^b$Dyudina et al. [2010], $^c$Evans et al. [1983], $^d$Little et
al. [1999], $^e$Lanzerotti et al. [1996], $^f$Rakov and Uman [2003],
and $^g$Christian et al. [1989].}\label{ftable2}
\begin{center}
\begin{tabular}{|c|c|c|c|} % this is to center (c) all numbers
% with vertical lines in between, use l or r for flush left or right
\hline % this makes a horizontal line
Characteristic & Saturn & Jupiter & Earth\\
\hline
Storm duration & days/months & days & tens of minutes/hours\\
Storm size & $^a$ 2000-3000~km & $^d <$1500~km & $^f$ 25~km\\
Illuminated cloud region & $^b$ 100~km & $^d$ 45-80~km & $^g$ 10~km\\
Optical flash energy & $^b$ 10$^9$~J & $^d$ 10$^9$~J & $^f$ 10$^6$~J\\
Total flash energy & 10$^{12}$~J? & 10$^{12}$~J? & $^f$ 10$^9$~J\\
Stroke/flash duration & $^c <140~\mu s$?/70~ms & $^e$ 500~$\mu$s?/? & $^f$ 70~$\mu$s/300~ms\\
\hline
\end{tabular}
\end{center}
\end{table}
For Saturn and Jupiter the illuminated cloud regions seen from above
are the half width at half maximum (HWHM) of the flash brightness
measured from the flash center. The larger size for Saturn is due to
the greater depth of the lightning source below the cloud tops
[Dyudina et al., 2010]. For Earth, illuminated cloud regions have a
typical size of 10~km, but can be up to 60~km for large storms
[Christian et al., 1989]. The optical energies for Jovian and
Saturnian flashes are similar ($\sim 10^9$~J), and about 3 orders of
magnitude larger compared to a normal terrestrial flash (not a
superbolt). Using an optical efficiency of $\sim 0.1$\% [Borucki and
McKay, 1987], we estimate that the total energy of a lightning flash
on Jupiter and Saturn is $\sim 10^{12}$~J. In Table~\ref{ftable2} we
distinguish between stroke and flash duration since it is very
likely that flashes on Jupiter and Saturn consist of sub--strokes.
The Galileo Probe has measured waveforms of Jovian strokes
[Lanzerotti et al., 1996], but it is unclear if they form a longer
flash. Jovian lightning has no high frequency radio component
detectable from space with which this question could be answered.
For Saturn we probably have not yet resolved the sub--structure of
the flashes.

\section{Summary and conclusions}

The Cassini RPWS instrument has continuously monitored the radio
emissions of Saturn lightning during the last six years. This paper
shows that studying Saturn lightning has become a
multi--instrumental task, and other instruments have provided
important new clues and additional information. Besides RPWS it
involves the Cassini instruments ISS, VIMS, CIRS, as well as
ground--based optical observations and measurements by giant radio
telescopes. This fruitful cooperation should be continued in future
and last at least until the end of the Cassini mission in 2017.

\section*{Acknowledgement}

G. F. is funded by a grant (P21295--N16) from the Austrian Science
Fund (FWF).

\section*{References}
\everypar={\hangindent=1truecm \hangafter=1}

% Use the references.tex file to look for commonly used
% planetary radio science references, and then simply copy and
% paste it. Our reference list is quite long, in case you
% don't find the reference, please write it in the same way
% as given the example references below.

Baines, K. H., M. L. Delitsky, T. W. Momary, R. H. Browan, B. J.
Buratti, R. N. Clark, and P. D. Nicholson, Storm clouds on Saturn:
Lightning--induced chemistry and associated materials consistent
with Cassini/VIMS spectra, \textsl{Planet. Space Sci.}, \textbf{57},
1650--1658, 2009.

Bar--Nun, A., Thunderstorms on Jupiter, \textsl{Icarus},
\textbf{24}, 86--94, 1975.

Borucki, W. J., and C. P. McKay, Optical efficiencies of lightning
in planetary atmospheres, \textsl{Nature}, \textbf{328}, 509--510,
1987.

Christian, H. J., R. J. Blakeslee, and S. J. Goodman, The detection
of lightning from geostationary orbit, \textsl{J. Geophys. Res.},
\textbf{94}, D11, 13329--13337, 1989.

%Brown, R. H., et al. (21 co--authors), The Cassini Visual and
%Infrared Mapping Spectrometer (VIMS) investigation, \textsl{Space
%Sci. Rev.}, \textbf{115}, 111--168, 2004.

Delcroix, M., and G. Fischer, Contribution of amateur observations
to Saturn storm studies, EPSC 2010--132 Abstracts Vol. 5, European
Planetary Science Congress, Rome, Italy, 2010.

Dyudina, U. A., A. P. Ingersoll, S. P. Ewald, C. C. Porco, G.
Fischer, W. S. Kurth, M. D. Desch, A. Del Genio, J. Barbara, and J.
Ferrier, Lightning storms on Saturn observed by Cassini ISS and RPWS
during 2004--2006, \textsl{Icarus}, \textbf{190}, 545--555, 2007.

Dyudina, U. A., A. P. Ingersoll, S. P. Ewald, C. C. Porco, G.
Fischer, W. S. Kurth, and R. A. West, Detection of visible lightning
on Saturn, \textsl{Geophys. Res. Lett.}, \textbf{37}, L09205, 2010.

Evans, D. R., J. H. Romig, and J. W. Warwick, Saturn's Electrostatic
Discharges: Properties and theoretical considerations,
\textsl{Icarus}, \textbf{54}, 267--279, 1983.

%Flasar, F. M., et al. (44 co--authors), Exploring the Saturn system
%in the thermal infrared: The Composite Infrared Spectrometer,
%\textsl{Space Sci. Rev.}, \textbf{115}, 169--297, 2004.

Fischer, G., M. D. Desch, P. Zarka, M. L. Kaiser, D. A. Gurnett, W.
S. Kurth, W. Macher, H. O. Rucker, A. Lecacheux, W. M. Farrell, and
B. Cecconi, Saturn lightning recorded by Cassini/RPWS in 2004,
\textsl{Icarus}, \textbf{183}, 135--152, 2006.

Fischer, G., W. S. Kurth, U. A. Dyudina, M. L. Kaiser, P. Zarka, A.
Lecacheux, A. P. Ingersoll, and D. A. Gurnett, Analysis of a giant
lightning storm on Saturn, \textsl{Icarus}, \textbf{190}, 528--544,
2007.

%Fischer, G., D. A. Gurnett, W. S. Kurth, F. Akalin, P. Zarka, U. A.
%Dyudina, W. M. Farrell, and M. L. Kaiser, Atmospheric electricity at
%Saturn, \textsl{Space Sci. Rev.}, \textbf{137}, 271--285, 2008.

%Gurnett, D. A., et al. (29 co--authors), The Cassini Radio and
%Plasma Wave investigation, \textsl{Space Sci. Rev.}, \textbf{114},
%1, 395--463, 2004.

Grie{\ss}meier, J.--M., P. Zarka, A. A. Konovalenko, G. Fischer, V.
V. Zakharenko, B. Ryabov, D. Vavriv, V. Ryabov, H. O. Rucker, and
the Radio--Exopla Collaboration, Ground--based study of Saturn
lightning, in \textsl{Planetary Radio Emissions VII}, edited by H.
O. Rucker, W. S. Kurth, P. Louarn, and G. Fischer, Austrian Academy
of Sciences Press, Vienna, this issue, 2011.

Hueso, R., and A. Sanchez--Lavega, A three--dimensional model of
moist convection for the giant planets. II. Saturn's water and
ammonia moist convective storms, \textsl{Icarus}, \textbf{172},
255--271, 2004.

Hueso, R., J. Legarreta, S. P\'{e}rez--Hoyos, J. F. Rojas, A.
S\'{a}nchez--Lavega, and A. Morgado, The international outer planets
watch atmospheres node database of giant--planet images,
\textsl{Planet. Space Sci.}, \textbf{58}, 1152--1159, 2010.

Konovalenko, A. A., A. Lecacheux, H. O. Rucker, G. Fischer, E. P.
Abranin, N. N. Kalinichenko, I. S. Falkovch, and K. M. Sidorchuk,
Ground--based decameter wavelength observations of Saturn
electrostatic discharges, European Planetary Science Congress,
EPSC2006--A--00229, Berlin, 2006.

Lanzerotti, L. J., K. Rinnert, G. Dehmel, F. O. Gliem, E. P. Krider,
M. A. Uman, and J. Bach, Radio frequency signals in Jupiter's
atmosphere, \textsl{Science}, \textbf{272}, 858--860, 1996.

Little, B., C. D. Anger, A. P. Ingersoll, A. R. Vasavada, D. A.
Senske, H. H. Breneman, W. J. Borucki, and the Galileo SSI Team,
Galileo images of lightning on Jupiter, \textsl{Icarus},
\textbf{142}, 306--323, 1999.

%Porco, C. C., et al. (19 co--authors), Cassini Imaging Science:
%Instrument characteristics and anticipated scientific investigations
%at Saturn, \textsl{Space Sci. Rev.}, \textbf{115}, 1--4, 363--497,
%2004.

Porco, C. C., et al. (34 co--authors), Cassini imaging science:
Initial results on Saturn's atmosphere, \textsl{Science},
\textbf{307}, 1243--1247, 2005.

Rakov, V. A., and M. A. Uman, Lightning: Physics and Effects,
Cambridge Univ. Press, Cambridge, 2003.

Sanchez--Lavega, A., F. Colas, J. Lecacheux, P. Laques, D. Parker,
and I. Miyazaki, The Great White Spot and disturbances in Saturn's
equatorial atmosphere during 1990, \textsl{Nature}, \textbf{353},
397--401, 1991.

Stoker, C. R., Moist convection -- A mechanism for producing the
vertical structure of the Jovian equatorial plumes, \textsl{Icarus},
\textbf{76}, 106--125, 1986.

Warwick, J. W., J. B. Pearce, D. R. Evans, T. D. Carr, J. J.
Schauble, J. K. Alexander, M. L. Kaiser, M. D. Desch, B. M.
Pedersen, A. Lecacheux, G. Daigne, A. Boischot, and C. H. Barrow,
Planetary Radio Astronomy observations from Voyager~1 near Saturn,
\textsl{Science}, \textbf{212}, 239--243, 1981.

Zarka, P., W. M. Farrell, G. Fischer, and A. Konovalenko,
Ground--based and space--based radio observations of planetary
lightning, \textsl{Space Sci. Rev.}, \textbf{137}, 257--269, 2008.

\end{document}